
%
\def\unlockat{\catcode`\@=11}
\def\lockat{\catcode`\@=12}
\unlockat
\def\d@f@ult{} \newif\ifamsfonts \newif\ifafour
\def\m@ssage{\immediate\write16}  \m@ssage{}
\d@f@ult\amsfontstrue
%
\d@f@ult\afourtrue
%
\nonstopmode
%
%

\font\twelverm=cmr12
\font\ninerm=cmr9
\font\sixrm=cmr6
\font\fourteenbf=cmbx12 scaled\magstep1
\font\twelvebf=cmbx12
\font\ninebf=cmbx9
\font\sixbf=cmbx6
\font\fourteeni=cmmi12 scaled\magstep1      \skewchar\fourteeni='177
\font\twelvei=cmmi12                        \skewchar\twelvei='177
\font\ninei=cmmi9                           \skewchar\ninei='177
\font\sixi=cmmi6                            \skewchar\sixi='177
\font\fourteensy=cmsy10 scaled\magstep2     \skewchar\fourteensy='60
\font\twelvesy=cmsy10 scaled\magstep1       \skewchar\twelvesy='60
\font\ninesy=cmsy9                          \skewchar\ninesy='60
\font\sixsy=cmsy6                           \skewchar\sixsy='60
\font\fourteenex=cmex10 scaled\magstep2
\font\twelveex=cmex10 scaled\magstep1

\ifamsfonts
   \font\ninex=cmex9
   
   \font\sixex=cmex7 at 6pt
   
\else
   \font\ninex=cmex10 at 9pt
   
   \font\sixex=cmex10 at 6pt
   
\fi
\font\fourteensl=cmsl10 scaled\magstep2
\font\twelvesl=cmsl10 scaled\magstep1

\font\sevensl=cmsl10 at 7pt
\font\sixsl=cmsl10 at 6pt

\font\fourteenit=cmti12 scaled\magstep1
\font\twelveit=cmti12

\font\fourteentt=cmtt12 scaled\magstep1
\font\twelvett=cmtt12
\font\fourteencp=cmcsc10 scaled\magstep2
\font\twelvecp=cmcsc10 scaled\magstep1

\ifamsfonts
   
\else
   
\fi
\newfam\cpfam
\font\fourteenss=cmss12 scaled\magstep1
\font\twelvess=cmss12
\font\tenss=cmss10
\font\niness=cmss9

\font\sevenss=cmss8 at 7pt
\font\sixss=cmss8 at 6pt
\newfam\ssfam
\newfam\msafam \newfam\msbfam \newfam\eufam
\ifamsfonts
 \font\fourteenmsa=msam10 scaled\magstep2
 \font\twelvemsa=msam10 scaled\magstep1
 \font\tenmsa=msam10
 \font\ninemsa=msam9
 \font\sevenmsa=msam7
 \font\sixmsa=msam6
 \font\fourteenmsb=msbm10 scaled\magstep2
 \font\twelvemsb=msbm10 scaled\magstep1
 \font\tenmsb=msbm10
 \font\ninemsb=msbm9
 \font\sevenmsb=msbm7
 \font\sixmsb=msbm6
 \font\fourteeneu=eufm10 scaled\magstep2
 \font\twelveeu=eufm10 scaled\magstep1
 \font\teneu=eufm10
 \font\nineeu=eufm9
 
 \font\seveneu=eufm7
 \font\sixeu=eufm6
 \def\hexnumber@#1{\ifnum#1<10 \number#1\else
  \ifnum#1=10 A\else\ifnum#1=11 B\else\ifnum#1=12 C\else
  \ifnum#1=13 D\else\ifnum#1=14 E\else\ifnum#1=15 F\fi\fi\fi\fi\fi\fi\fi}
 \def\hexmsa{\hexnumber@\msafam}
 \def\hexmsb{\hexnumber@\msbfam} 
\fi
\newdimen\b@gheight             \b@gheight=12pt
\newcount\f@ntkey               \f@ntkey=0
\def\f@m{\afterassignment\samef@nt\f@ntkey=}
\def\samef@nt{\fam=\f@ntkey \the\textfont\f@ntkey\relax}
\def\rm{\f@m0 }
\def\mit{\f@m1 }
\def\cal{\f@m2 }
\def\it{\f@m\itfam}
\def\sl{\f@m\slfam}
\def\bf{\f@m\bffam}
\def\tt{\f@m\ttfam}
\def\caps{\f@m\cpfam}
\def\ssf{\f@m\ssfam}
\ifamsfonts
 \def\msa{\f@m\msafam}
 \def\msb{\f@m\msbfam} 
 \def\eu{\f@m\eufam}
\else
  \let\eu=\bf
\fi
\def\fourteenpoint{\relax
    \textfont0=\fourteencp          \scriptfont0=\tenrm
      \scriptscriptfont0=\sevenrm
    \textfont1=\fourteeni           \scriptfont1=\teni
      \scriptscriptfont1=\seveni
    \textfont2=\fourteensy          \scriptfont2=\tensy
      \scriptscriptfont2=\sevensy
    \textfont3=\fourteenex          \scriptfont3=\twelveex
      \scriptscriptfont3=\tenex
    \textfont\itfam=\fourteenit     \scriptfont\itfam=\tenit
    \textfont\slfam=\fourteensl     \scriptfont\slfam=\tensl
      \scriptscriptfont\slfam=\sevensl
    \textfont\bffam=\fourteenbf     \scriptfont\bffam=\tenbf
      \scriptscriptfont\bffam=\sevenbf
    \textfont\ttfam=\fourteentt
    \textfont\cpfam=\fourteencp
    \textfont\ssfam=\fourteenss     \scriptfont\ssfam=\tenss
      \scriptscriptfont\ssfam=\sevenss
    \ifamsfonts
       \textfont\msafam=\fourteenmsa     \scriptfont\msafam=\tenmsa
         \scriptscriptfont\msafam=\sevenmsa
       \textfont\msbfam=\fourteenmsb     \scriptfont\msbfam=\tenmsb
         \scriptscriptfont\msbfam=\sevenmsb
       \textfont\eufam=\fourteeneu     \scriptfont\eufam=\teneu
         \scriptscriptfont\eufam=\seveneu \fi
    \samef@nt
    \b@gheight=14pt
    \setbox\strutbox=\hbox{\vrule height 0.85\b@gheight
                                depth 0.35\b@gheight width\z@ }}
\def\twelvepoint{\relax
    \textfont0=\twelverm          \scriptfont0=\ninerm
      \scriptscriptfont0=\sixrm
    \textfont1=\twelvei           \scriptfont1=\ninei
      \scriptscriptfont1=\sixi
    \textfont2=\twelvesy           \scriptfont2=\ninesy
      \scriptscriptfont2=\sixsy
    \textfont3=\twelveex          \scriptfont3=\ninex
      \scriptscriptfont3=\sixex
    \textfont\itfam=\twelveit    
    \textfont\slfam=\twelvesl    
      \scriptscriptfont\slfam=\sixsl
    \textfont\bffam=\twelvebf     \scriptfont\bffam=\ninebf
      \scriptscriptfont\bffam=\sixbf
    \textfont\ttfam=\twelvett
    \textfont\cpfam=\twelvecp
    \textfont\ssfam=\twelvess     \scriptfont\ssfam=\niness
      \scriptscriptfont\ssfam=\sixss
    \ifamsfonts
       \textfont\msafam=\twelvemsa     \scriptfont\msafam=\ninemsa
         \scriptscriptfont\msafam=\sixmsa
       \textfont\msbfam=\twelvemsb     \scriptfont\msbfam=\ninemsb
         \scriptscriptfont\msbfam=\sixmsb
       \textfont\eufam=\twelveeu     \scriptfont\eufam=\nineeu
         \scriptscriptfont\eufam=\sixeu \fi
    \samef@nt
    \b@gheight=12pt
    \setbox\strutbox=\hbox{\vrule height 0.85\b@gheight
                                depth 0.35\b@gheight width\z@ }}
\twelvepoint
%
%
\baselineskip = 15pt plus 0.2pt minus 0.1pt 
\lineskip = 1.5pt plus 0.1pt minus 0.1pt
\lineskiplimit = 1.5pt
\parskip = 6pt plus 2pt minus 1pt
\interlinepenalty=50
\interfootnotelinepenalty=5000
\predisplaypenalty=9000
\postdisplaypenalty=500
\hfuzz=1pt
\vfuzz=0.2pt
\dimen\footins=24 truecm 
\ifafour
 \hsize=16cm \vsize=22cm
\else
 \hsize=6.5in \vsize=9in
\fi
%
%
\skip\footins=\medskipamount
\newcount\fnotenumber
\def\clearfnotenumber{\fnotenumber=0} \clearfnotenumber
\def\fnote{\global\advance\fnotenumber by1 \generatefootsymbol
 \footnote{$^{\footsymbol}$}}
\def\fd@f#1 {\xdef\footsymbol{\mathchar"#1 }}
\def\generatefootsymbol{\iffrontpage\ifcase\fnotenumber
\or \fd@f 279 \or \fd@f 27A \or \fd@f 278 \or \fd@f 27B
\else  \fd@f 13F \fi
\else\xdef\footsymbol{\the\fnotenumber}\fi}
%
%
\newcount\secnumber \newcount\appnumber
\def\clearappnumber{\appnumber=64} \def\clearsecnumber{\secnumber=0}
\clearsecnumber \clearappnumber
\newif\ifs@c 
\newif\ifs@cd 
\s@cdtrue 
\def\unsectioned{\s@cdfalse\let\section=\subsection}
\newskip\sectionskip         \sectionskip=\medskipamount
\newskip\headskip            \headskip=8pt plus 3pt minus 3pt
\newdimen\sectionminspace    \sectionminspace=10pc
\def\Titlestyle#1{\par\begingroup \interlinepenalty=9999
     \leftskip=0.02\hsize plus 0.23\hsize minus 0.02\hsize
     \rightskip=\leftskip \parfillskip=0pt
     \advance\baselineskip by 0.5\baselineskip
     \hyphenpenalty=9000 \exhyphenpenalty=9000
     \tolerance=9999 \pretolerance=9000
     \spaceskip=0.333em \xspaceskip=0.5em
     \fourteenpoint
  \noindent #1\par\endgroup }
\def\titlestyle#1{\par\begingroup \interlinepenalty=9999
     \leftskip=0.02\hsize plus 0.23\hsize minus 0.02\hsize
     \rightskip=\leftskip \parfillskip=0pt
     \hyphenpenalty=9000 \exhyphenpenalty=9000
     \tolerance=9999 \pretolerance=9000
     \spaceskip=0.333em \xspaceskip=0.5em
     \fourteenpoint
   \noindent #1\par\endgroup }
\def\spacecheck#1{\dimen@=\pagegoal\advance\dimen@ by -\pagetotal
   \ifdim\dimen@<#1 \ifdim\dimen@>0pt \vfil\break \fi\fi}
\def\section#1{\cleareqnumber \s@ctrue \global\advance\secnumber by1
   \par \ifnum\the\lastpenalty=30000\else
   \penalty-200\vskip\sectionskip \spacecheck\sectionminspace\fi
   \noindent {\caps\enspace\S\the\secnumber\quad #1}\par
   \nobreak\vskip\headskip \penalty 30000 }
\def\undertext#1{\vtop{\hbox{#1}\kern 1pt \hrule}}
\def\subsection#1{\par
   \ifnum\the\lastpenalty=30000\else \penalty-100\smallskip
   \spacecheck\sectionminspace\fi
   \noindent\undertext{#1}\enspace \vadjust{\penalty5000}}

\def\appendix#1{\cleareqnumber \s@cfalse \global\advance\appnumber by1
   \par \ifnum\the\lastpenalty=30000\else
   \penalty-200\vskip\sectionskip \spacecheck\sectionminspace\fi
   \noindent {\caps\enspace Appendix \char\the\appnumber\quad #1}\par
   \nobreak\vskip\headskip \penalty 30000 }
\def\ack{\par\penalty-100\medskip \spacecheck\sectionminspace
   \line{\fourteencp\hfil ACKNOWLEDGEMENTS\hfil}%
\nobreak\vskip\headskip }
\def\refs{\begingroup \par\penalty-100\medskip \spacecheck\sectionminspace
   \line{\fourteencp\hfil REFERENCES\hfil}%
\nobreak\vskip\headskip \frenchspacing }
\def\endrefs{\par\endgroup}
%
%
\newif\iffrontpage \frontpagefalse
\headline={\hfil}
\footline={\iffrontpage\hfil\else \hss\twelverm
-- \folio\ --\hss \fi }
%
%
\newskip\frontpageskip \frontpageskip=12pt plus .5fil minus 2pt
\def\titlepage{\global\frontpagetrue\hrule height\z@ \relax
               \pubblock\relax }
\def\endtitlepage{\vfil\break\clearfnotenumber\frontpagefalse}
\def\title#1{\vskip\frontpageskip\Titlestyle{\caps #1}\vskip3\headskip}
\def\author#1{\vskip.5\frontpageskip\titlestyle{\caps #1}\nobreak}
\def\and{\par\kern 5pt \centerline{\sl and}}

\def\address#1{\par\kern 5pt\titlestyle{\it #1}}
\def\andaddress{\par\kern 5pt \centerline{\sl and} \address}

\def\abstract#1{\par\dimen@=\prevdepth \hrule height\z@ \prevdepth=\dimen@
   \vskip\frontpageskip\spacecheck\sectionminspace
   \centerline{\fourteencp ABSTRACT}\vskip\headskip
   {\noindent #1}}

\def\email#1{\fnote{\tentt e-mail: #1\hfill}}

%
%

%

%

%

%

%
\def\UB{\address{%
   Departament d'Estructura i Constituents de la Mat{\`e}ria\break
   Universitat de Barcelona, Diagonal 647\break
   E-08028 BARCELONA}}
%
%
\newcount\refnumber \def\clearrefnumber{\refnumber=0}  \clearrefnumber
\newwrite\R@fs                              
\immediate\openout\R@fs=\jobname.refs 
\def\closerefs{\immediate\closeout\R@fs} 
\def\refsout{\closerefs\refs
\unlockat
\input\jobname.refs
\lockat
\endrefs}
\def\refitem#1{\item{{\bf #1}}}
\def\ifundefined#1{\expandafter\ifx\csname#1\endcsname\relax}
\def\[#1]{\ifundefined{#1R@FNO}%
\global\advance\refnumber by1%
\expandafter\xdef\csname#1R@FNO\endcsname{[\the\refnumber]}%
\immediate\write\R@fs{\noexpand\refitem{\csname#1R@FNO\endcsname}%
\noexpand\csname#1R@F\endcsname}\fi{\bf \csname#1R@FNO\endcsname}}
\def\refdef[#1]#2{\expandafter\gdef\csname#1R@F\endcsname{{#2}}}
%
%
\newcount\eqnumber \def\cleareqnumber{\eqnumber=0}
\newif\ifal@gn \al@gnfalse  
\def\veqnalign#1{\al@gntrue \vbox{\eqalignno{#1}} \al@gnfalse}
\def\eqnalign#1{\al@gntrue \eqalignno{#1} \al@gnfalse}
\def\(#1){\relax%
\ifundefined{#1@Q}
 \global\advance\eqnumber by1
 \ifs@cd
  \ifs@c
   \expandafter\xdef\csname#1@Q\endcsname{{%
\noexpand\rm(\the\secnumber .\the\eqnumber)}}
  \else
   \expandafter\xdef\csname#1@Q\endcsname{{%
\noexpand\rm(\char\the\appnumber .\the\eqnumber)}}
  \fi
 \else
  \expandafter\xdef\csname#1@Q\endcsname{{\noexpand\rm(\the\eqnumber)}}
 \fi
 \ifal@gn
    & \csname#1@Q\endcsname
 \else
    \eqno \csname#1@Q\endcsname
 \fi
\else%
\csname#1@Q\endcsname\fi\global\let\@Q=\relax}
%
%
\newif\ifm@thstyle \m@thstylefalse
\def\mathstyle{\m@thstyletrue}
\def\proclaim#1#2\par{\smallbreak\begingroup
\advance\baselineskip by -0.25\baselineskip%
\advance\belowdisplayskip by -0.35\belowdisplayskip%
\advance\abovedisplayskip by -0.35\abovedisplayskip%
    \noindent{\caps#1.\enspace}{#2}\par\endgroup%
\smallbreak}
\def\m@kem@th<#1>#2#3{%
\ifm@thstyle \global\advance\eqnumber by1
 \ifs@cd
  \ifs@c
   \expandafter\xdef\csname#1\endcsname{{%
\noexpand #2\ \the\secnumber .\the\eqnumber}}
  \else
   \expandafter\xdef\csname#1\endcsname{{%
\noexpand #2\ \char\the\appnumber .\the\eqnumber}}
  \fi
 \else
  \expandafter\xdef\csname#1\endcsname{{\noexpand #2\ \the\eqnumber}}
 \fi
 \proclaim{\csname#1\endcsname}{#3}
\else
 \proclaim{#2}{#3}
\fi}
\def\Thm<#1>#2{\m@kem@th<#1M@TH>{Theorem}{\sl#2}}
\def\Prop<#1>#2{\m@kem@th<#1M@TH>{Proposition}{\sl#2}}
\def\Def<#1>#2{\m@kem@th<#1M@TH>{Definition}{\rm#2}}
\def\Lem<#1>#2{\m@kem@th<#1M@TH>{Lemma}{\sl#2}}
\def\Cor<#1>#2{\m@kem@th<#1M@TH>{Corollary}{\sl#2}}
\def\Conj<#1>#2{\m@kem@th<#1M@TH>{Conjecture}{\sl#2}}
\def\Rmk<#1>#2{\m@kem@th<#1M@TH>{Remark}{\rm#2}}
\def\Exm<#1>#2{\m@kem@th<#1M@TH>{Example}{\rm#2}}
\def\Qry<#1>#2{\m@kem@th<#1M@TH>{Query}{\it#2}}
%
%

%
\def\<#1>{\csname#1M@TH\endcsname}
%
%
\def\ref#1{{\bf [#1]}}
\def\ie{{\it i.e.\/}}
%
%

\def\lapprox{\hbox{\lower3pt\hbox{$\buildrel<\over\sim$}}}
\def\gapprox{\hbox{\lower3pt\hbox{$\buildrel<\over\sim$}}}
\def\quotient#1#2{#1/\lower0pt\hbox{${#2}$}}
\def\fr#1/#2{\mathord{\hbox{${#1}\over{#2}$}}}
\ifamsfonts
 \mathchardef\empty="0\hexmsb3F 
 \mathchardef\lsemidir="2\hexmsb6E 
 \mathchardef\rsemidir="2\hexmsb6F 
\else
 \let\empty=\emptyset
 \def\lsemidir{\mathbin{\hbox{\hskip2pt\vrule height 5.7pt depth -.3pt
    width .25pt\hskip-2pt$\times$}}}
 \def\rsemidir{\mathbin{\hbox{$\times$\hskip-2pt\vrule height 5.7pt
    depth -.3pt width .25pt\hskip2pt}}}
\fi
%
%

%
%
%
%
\def\underrightarrow#1{\vtop{\ialign{##\crcr
      $\hfil\displaystyle{#1}\hfil$\crcr
      \noalign{\kern-\p@\nointerlineskip}
      \rightarrowfill\crcr}}} 
\def\underleftarrow#1{\vtop{\ialign{##\crcr
      $\hfil\displaystyle{#1}\hfil$\crcr
      \noalign{\kern-\p@\nointerlineskip}
      \leftarrowfill\crcr}}}  

%
%
%
%

\def\NPB#1#2#3{{\sl Nucl. Phys.} {\bf B#1} (#2) #3}
\def\NPBFS#1#2#3#4{{\sl Nucl. Phys.} {\bf B#2} [FS#1] (#3) #4}

\def\PRD#1#2#3{{\sl Phys. Rev.} {\bf D#1} (#2) #3}
\def\PLA#1#2#3{{\sl Phys. Lett.} {\bf #1A} (#2) #3}
\def\PLB#1#2#3{{\sl Phys. Lett.} {\bf #1B} (#2) #3}

\lockat


%
%

\def\fr#1/#2{\mathord{\hbox{${#1}\over{#2}$}}}

\def\ket|#1>{\mathord{\vert{#1}\rangle}}

\def\ope#1#2{{{#2}\over{\ifnum#1=1 {z-w} \else {(z-w)^{#1}}\fi}}}

\def\corr<#1>{\mathord{\langle #1 \rangle}}

%

%
%

\def\kink{\varphi_{\hskip-2pt{}_K}}
%

%
%
\refdef[MorseFeshbach]
{P. Morse and H. Feshbach, {Methods of Mathematical Physics},
McGraw-Hill Book Company, 1953.}
\refdef[Rajaraman]
{R. Rajaraman, {\sl Solitons and Instantons}, 
North-Holland Publishing Company, 1982.}
\refdef[NielsenOlesen]
{H.B. Nielsen and P. Olesen, \NPB{61}{1973}{45}.}
\refdef[Forster]
{D. F{\"o}rster, \NPB{81}{1974}{84}.}
\refdef[TurokMadea]
{N. Turok and K. Madea, \PLB{202}{1988}{376}.\hfill\break
{V. Silveira and M.D. Maia, \PLA{174}{1993}{280}.}\hfill\break
A.L. Larsen, \PLA{181}{1993}{369}.}
\refdef[Gregory]
{R. Gregory, \PRD{43}{1991}{520}.}
\refdef[Kaputsnikovetal]
{A.A. Kaputsnikov, A. Pashnev and A. Pichugin, \PRD{55}{1997}{2257}.}
\refdef[Arodz]
{H. Arod\'z, \NPB{450}{1995}{174}.}
\refdef[CarterGregory]
{B. Carter and R. Gregory, \PRD{51}{1995}{5839}.}
\refdef[Zia]
{R.K.P. Zia, \NPBFS{13}{251}{1985}{676}.}
\refdef[RamosRoca]
{E. Ramos and J. Roca, \PLB{366}{1996}{113}.}
\refdef[Diehletal]
{H.W. Diehl, D.M. Kroll and H. Wagner
{\sl Z. Physik} {\bf B36} (1980) 329.}
\refdef[ArodzWegrzyn]
{H. Arod\'z and P. W\c egrzyn, \PLB{291}{1992}{251}.}

%
\overfullrule=0pt
\def\pubblock{ \line{\hfil\rm UB-ECM-PF 97/06}
               \line{\hfil\tt hep-th/9705085}
               \line{\hfil\rm May 1997}
               \line{\hfil\rm (January 1998)}}   
\titlepage

\title{A Remark on the Effective Description of Topological Defects}

\author{Jordi Par{\'\i}s\email{paris@ecm.ub.es}
        and Jaume Roca\email{roca@ecm.ub.es}}
\UB

%

\abstract{We subject the methodology used to derive the effective
dynamics of topological defects to a critical reappraisal, using
the two-dimensional kink as an illustrative example. Special care
is taken on how the zero modes should be handled in order to avoid
overcounting of degrees of freedom. This is an issue that has been
overlooked in many recent contributions on the derivation of domain
wall effective actions. We show that, unless such redundancy is 
completely removed by means of a sort of gauge-fixing, the expression
obtained for the effective action will not be consistent. We 
readdress some earlier calculations over the existence of curvature
corrections in the light of the previous discussion and briefly 
comment on the application of this method to higher dimensional
topological defects.}

\endtitlepage

\section {Introduction}

Whenever we have a field theory with a set of vacua given by a non-connected
space there is the possibility of having different regions in space
living on different vacuum sectors. 
Two such regions will meet at what is generally named a topological defect,
\ie\ a thin hypersurface where the field rapidly evolves from
one vacuum to the other.

Field configurations of this kind will be stable against decay into
any of the true vacua provided that all such space regions are of an 
infinite volume, as in this case one would need an infinite amount of
energy to wipe the domain wall(s) off the game.
As a consequence, the space of finite energy field configurations will
itself also
be composed of a number of disconnected sectors, each of them being
characterized by the asymptotics of the relevant field.

The simplest example giving rise to this behavior is provided by the
two-dimensional spontaneously broken $\varphi^4$ theory:
$$
S_0[\varphi]={1\over{\lambda^2}}\int d^2x\;
\left[{1\over2}\partial_\mu\varphi\partial^\mu\varphi
-{{m^2}\over4}(\varphi^2-1)^2\right].
\(action)
$$

The set of vacua for this theory, given by the field 
configurations $\varphi=\pm1$, 
allows for the existence of a so-called topological sector in field 
space, characterized by the asymptotic behavior
$$
\varphi\rightarrow\pm1\quad{\rm for}\quad x\rightarrow\pm\infty,
\(boundcond)
$$
respectively.

The lowest energy solution in this sector obeys the equation
$$
-\partial^2_x\kink + m^2(\kink^2-1)\;\kink=0,
\(kinkeq)
$$
and it is explicitly given by
$$
\kink(x)={\rm th}\left({ m\over{\sqrt2}}\;x\right).
\(kink)
$$
It describes a localized finite-energy configuration, the {\it kink},
with a typical width of the order $1/m$ and sitting at rest on
the origin.
With a mass $M_K\sim m/\lambda^2$,
it is clearly a non-perturbative solution of the equations of motion.
We can say that the location of the kink's core is what can be 
identified as the topological defect separating the two vacua.

Small perturbations around the static kink are governed, to the lowest
order, by the equation
$$
-\partial^2_t\delta\varphi(t,x)
=\left[-\partial^2_x+m^2(3\kink^2(x)-1)\right]
\delta\varphi(t,x).
\(perturbeq)
$$

The spectrum of the operator on the right-hand-side consists of
two ``bound states'' and a continuum of ``scattering states'' labeled
by their momentum \[MorseFeshbach]\[Rajaraman].

The lowest mode $\delta\varphi_0=\kink'$ is actually a 
zero energy perturbation. It
denotes the presence of a flat direction in the potential,
corresponding to space translations of the kink.

Non-zero modes have a typical scale of the order $m$.
This means that, for perturbations such that $\delta E<<m$, these
modes will scarcely be excited and, in this situation, an effective 
description of the system can naturally be made by focussing on the
dynamics of the zero mode alone, {\it i.e.} in terms of the trajectory
of the kink's center of mass.

This sort of problem was originally addressed by Nielsen and Olesen
\[NielsenOlesen] who realized that the dynamics of vortices can be
approximately described by a string model. Soon after, F{\"o}rster
\[Forster] introduced a covariant method to obtain the effective dynamics
of these vortex lines, giving rise to the Nambu-Goto action in the
zero-width limit.

The outline of the method, as one would apply it to the above system,
is roughly the following. One should consider the degrees of freedom of
the original field $\varphi$, in the kink sector, and split them
between the massless and massive modes:
$$
\varphi\quad\rightarrow\quad(x^\mu(s),\phi),
\(splitting)
$$
where $x^\mu(s)$ will describe in a covariant way the dynamics of the
zero mode, \ie\ the space-time location of the kink.
In order to find an effective action for this zero mode one should
``integrate'' over the field $\phi$. In a classical regime this
amounts to solving the equation of motion
$$
{\partial S\over\partial\phi}[x,\phi]=0,\quad\Rightarrow\quad
\phi=\phi[x(s)],
\(genphieq)
$$
leaving $x(s)$ as a sort of background field at this stage.
The effective action will then be obtained by the substitution of 
this solution back into the action:
$$
S_{\rm eff}[x]=S[x,\phi[x]].
\(effaction)
$$

It is simple to show that any solution $x^*(s)$ derived from 
$S_{\rm eff}[x]$ gives rise, through the assignment $\phi^*=\phi[x^*]$
and the change inverse to \(splitting), to a solution $\varphi^*$ 
of the original equations of motion obtained from $S[\varphi]$.

Clearly, the lowest order contribution in $S_{\rm eff}$ should be just
a free particle term with a mass given by the kink's mass $M_K$.
It might be, however, that higher order corrections could also be
present. Such corrections are expected to depend on the
geometrical invariants associated with the embedding $x(s)$, \ie\ the
world-line's curvature in this case.
These corrections would give rise to new solutions of the equations of
motion beyond the free kink solution. These would no longer be
lowest-energy solutions and the curvature terms
would generally produce a non-trivial evolution for $x(s)$.
This non-trivial behavior for the zero mode should be regarded as the
result of the dispute of this energy excess between the zero
and the excited modes.

In recent years, and mainly within the context of domain walls,
the evaluation of curvature corrections to the basic world-volume term
has attracted the attention of several authors  
\[TurokMadea]\[Gregory]\[Kaputsnikovetal],
giving rise to some controversy over the way how these corrections
should be computed.%
\fnote{A slightly different approach, which will not be touched upon 
here, focusses not on the effective action but directly on the 
obtention of the equations of motion governing the dynamics of the 
topological defect (see, for example, ref.\ \[Arodz]).}

We make in this paper a close look examination of the steps involved
in the standard application of this method for the computation of the
effective action.  Our main point is that a proper handling of degrees
of freedom requires to supplement the splitting process \(splitting)
with a sort of gauge-fixing condition that prevents overcounting of
zero modes.  We also show that leaving this redundancy unfixed gives
rise to unreliable results for the effective action.  For the sake of
simplicity, we use the two-dimensional kink as an illustrative
example, although the same conclusions apply for higher dimensional
topological defects as well.

The contents of the paper is organized as follows.
In \S2 we review the standard procedure of obtention of the effective
action by applying it to our simple example \(action).
\S3 contains a critical analysis of this method.  We show that an
unnoticed redundancy, a gauge symmetry in fact, slips in in the
standard treatment of the splitting procedure \(splitting), unless
special boundary conditions are assumed on the equations of motion.
We also derive the explicit form of these gauge transformations and
introduce a natural gauge-fixing for them.  A solution for the
properly gauge-fixed equation of motion \(genphieq) is trivially
found, implying that, in this new approach, the standard mass term
alone is not just an approximate solution for the effective action, to
leading order in $1/m$, but it is actually an exact solution.  We
devote \S4 to study further consequences of these new equations over
whether curvature contributions may or may not arise under some
circumstances.  We finish the paper with a few words on the
application of this method to general higher dimensional domain walls.

\section{Covariant approach to the effective dynamics}

We review in this section the standard use of the effective action
method by applying it to the two-dimensional model \(action).

The central idea \[Forster] is to explicitly bring into the problem, in a
covariant way, the variables describing the evolution of the defect.
One can do it by first making a change of space-time coordinates from 
Minkowski variables $x^\mu$ to a new set of adapted coordinates.
Part of these new coordinates parametrize the embedding of the
defect in space-time while the rest correspond to space-like
normal directions.

Doing this for the simple two-dimensional case, one will be changing the
space-time para\-metriz\-ation from $(x^0,x^1)$ to $(t,\rho)$ where
both sets of coordinates are related by%
\fnote{From now on $t$ does not necessarily stand for $x^0$ but will
represent an arbitrary para\-metrization of the world-line $x(t)$.
We will also set hereafter the coupling constant $\lambda$ in
\(action) to $1$ since it does not play any role for the on-going
discussions.}
$$
x^\mu = x^\mu(t) + \rho \;n^\mu(t).
\(change)
$$

Here $x^\mu(t)$ is meant to describe the world-line of the kink,
regarding it as a point particle, while $n^\mu(t)$ is a normalized
space-like vector, everywhere orthogonal to the unit tangent 
$v^\mu(t)$.
They obey the following relations
$$
v^\mu={1\over e}\;{{d x^\mu}\over{dt}},\quad\quad
n^\mu=\epsilon^{\mu\nu}v_\nu,
\(v&ndef)
$$ 
with $e=\sqrt{\dot x^2}$ and $\epsilon_{01}=+1$.

The geometrical interpretation of this change of variables is very
simple. Given an arbitrary space-time point $x^\mu $ we can get the
corresponding value for the parameter $t$ as the one labeling the 
point $x^\mu(t)$ on the world-line which is closest to $x^\mu$,
and the value for $\rho$ as the invariant distance between both 
points. 

The tangent and normal vectors $v$ and $n$ satisfy the Frenet
equations: 
$$
\eqnalign{
{1\over e}{{d v^\mu}\over{dt}}=&\;k\;n^\mu,
\cr
{1\over e}{{d n^\mu}\over{dt}}=&\;k\;v^\mu,
\(Freneteqns)}
$$
where $k$, the (signed) curvature, may assume either positive or
negative values. This is due to the definite orientation that we chose
for $n^\mu$ in \(v&ndef) which ensures that $(t,\rho)$ will 
coincide exactly with $(x^0,x^1)$ when $x^\mu(t)$ reduces to the 
world-line of a kink at rest on the origin.

It is clear that the change of variables \(change) cannot be
well-defined everywhere on space-time
unless the curvature $k$ vanishes for all $t$. 
As we can see from the Jacobian, $J=e\Delta$, with
$$
\Delta=1+\rho k,
\(Deltadef)
$$
we will be in trouble when trying to cover points $x^\mu $ farther 
away from the kink than the radius of curvature $1/k$.
The rationale for going ahead with these new variables relies on the
Lagrangian (and all other dynamical functions) being relevant only on
a small region surrounding the world-line $x^\mu(t)$.
This will indeed be the case for field configurations departing very
little (locally) from the kink solution, {\it i.e.} configurations for
which the typical curvature scale satisfies $|k|<<m$.
This is actually the regime we shall be concerned with. So, in this
situation, \(change) should be perfectly acceptable.

\medskip

One can now rewrite the action \(action) with the help of these new
coordinates,
$$
S_0[x,\phi]=\int dsd\rho\;\Delta\left[{1\over 2\Delta^2}(\partial_s\phi)^2
-{1\over 2}(\partial_\rho\phi)^2-{m^2\over 4}(\phi^2-1)^2\right],
\(curvedaction)
$$
where we have used the proper-time parametrization of $x^\mu(s)$
and have also defined $\phi(s,\rho)\equiv\varphi(x^\mu)$.

With this form of the action one may now consider the equation of
motion for $\phi$:
$$
{1\over\Delta}\;\partial_s\;{1\over\Delta}\;\partial_s\;\phi
\;-\;{1\over\Delta}\;\partial_\rho\;\Delta\;\partial_\rho\;\phi\;+\;
m^2(\phi^2-1)
\;\phi=0.
\(wrongeq)
$$

One can immediately check that the configuration 
$$
\phi(s,\rho)=\kink(\rho)
\(comovingkink)
$$
will not be in general a solution of equation \(wrongeq).
It describes a kink-like configuration whose center
is moving according to the world-line $x(s)$ and whose profile, in the
co-moving frame, is that of the static kink \(kink).
Inserting \(comovingkink) back into the action \(curvedaction) one would
get just the free particle action
$$
S_{\rm eff}[x]=-M_K\int ds,
\(freeaction)
$$
with a mass given by 
$$
M_K=\int d\rho(\kink')^2={2\sqrt2\over3}\;m.
\(kinkmass)
$$

The failure of \(comovingkink) to obey the $\phi$ equation of motion
is in a term which is proportional to the curvature $k$.
Of course, one cannot set $k$ to zero because $x^\mu(s)$ are being kept
as generic off-shell variables, but it leaves the door open for a
perturbative analysis in the regime of small values of $k$.

This sort of study, and its generalization to the case of higher
dimensional domain walls, has been performed by various authors 
\[Gregory]\[Kaputsnikovetal] in recent years.
The common idea was trying to get the explicit form of the 
curvature-dependent corrections to the free action \(freeaction) 
that would arise from corrections to \(comovingkink), which was taken
as the leading order solution for the $\phi$ equation of motion.

Let us sketch the general features of this procedure.
First, one assumes $k$ being of a typical scale $\Lambda$,
such that the ratio $\epsilon=\Lambda/m$ be actually very small.
This defines $\epsilon$ as the natural expansion parameter and 
it also confines the analysis to the regime of very slightly curved
$x(s)$ where the change of variables \(change) makes full sense.

A further, less intuitive, assumption is also issued. 
It forces the variation of $\phi$ along the $s$-direction to be of
order $\epsilon$ when compared to the variation along the $\rho$
direction.
This effectively makes the $s$-derivative term in \(wrongeq)
to be absent at leading order in the $\epsilon$ expansion and
it guarantees that the expansion
$$
\phi=\phi_{(0)}+\epsilon\;\phi_{(1)}+\ldots
\(phiexp)
$$
starts with the term $\phi_{(0)}=\kink(\rho)$.

It is also straightforward to check that $\phi_{(1)}$ should satisfy
the inhomogeneous equation
$$
-\partial^2_\rho\;\phi_{(1)}+m^2(3\kink^2-1)\;\phi_{(1)}=\kappa\;\kink',
\(phi1corr)
$$
with $k=\epsilon\;\kappa$. This implies that curvature corrections to 
\(freeaction) should be expected to arise
because of the $\phi_{(1)}$ contribution to the effective action.

Several authors have attempted to solve this equation with the help of
different additional assumptions but
we will not pursue this approach any further. We have only sketched 
it for later comparison with the analysis that we shall develop in the 
next section.
We address the interested reader to the original papers for further 
details about the above procedure.

\section{Proper handling of zero modes}

The main goal of this paper is to point out several
shortcomings of the methodology described in the previous section and
(hopefully) correct them.

First of all, we would like to recall that it is a standard assumption
in the literature to take the trajectory $x^\mu(t)$ of the
topological defect to be defined as
the locus of zero-field space-time points, {\it i.e.\ }those satisfying
$$
\varphi(x^\mu(t))=\phi(t,0)=0,\quad\forall t,
\(core)
$$ 
the so-called {\it core}.  This is a perfectly acceptable choice
provided that the field $\varphi$ is (locally) a small perturbation
from the standard kink solution because in this situation there will
be a single curve $x^\mu(t)$ satisfying $\varphi(x^\mu(t))=0$.
However there is, in our opinion, a missing ingredient in the
derivation of the equation of motion for $\phi$ in the way it has been
presented in \S2 and used in many recent contributions (see for example 
\[Gregory]\[Kaputsnikovetal]). This is because that derivation did not
fully take into consideration the constraint imposed by \(core). In
other words, when using the variational principle to derive the
equation of motion \(wrongeq) the fact that $\delta\phi(t,\rho)$
should be zero for $\rho=0$ was not taken into account.
The direct consequence of this omission is that equation \(wrongeq),
as it stands, is not completely correct. Indeed, being
$\delta\phi(t,0)=0$, the variational principle is still satisfied even
if the equation \(wrongeq) is not obeyed at $\rho=0$, implying that a
non-analytic behavior of $\phi$ should be allowed at those points. We
will devote the first part of this section to substantiate these
statements.

By looking at the previous section it is immediate to realize that
nowhere in the derivation of the equation of motion for $\phi$ was the
explicit relation \(core) between $\varphi$ and $x^\mu(t)$ ever used.
Consequently, we would have gotten the very same expressions for the
action \(curvedaction) and for the equation of motion \(wrongeq), if
that relation would have been different ({\it i.e.\ }if $x^\mu(t)$
were no longer the core) or even if there were no relation at all
between both objects ({\it i.e.\ }if $x^\mu(t)$ were an arbitrary
space-time curve, totally unrelated with $\varphi$).  This shows quite
clearly the fact that, although one might be assuming $x^\mu(t)$ to be
the core of $\varphi$, the subsequent steps followed to get the $\phi$
equation of motion did not take this restriction into account. In
fact, a gauge symmetry ---the ability to deform $x^\mu(t)$
arbitrarily--- has effectively slipped in.

In order to show what are the consequences of this gauge symmetry let us
first obtain its explicit expression. We can get it by considering a 
generic infinitesimal deformation of the curve $x(t)$
$$
\delta x^\mu(t)=\alpha(t)\; v^\mu(t)+\beta(t)\; n^\mu(t).
\(xtransform)
$$
Transformations generated by $\alpha$ correspond to world-line
reparametrizations whereas $\beta$ will be associated with the actual
deformations of its embedding in space-time.
Symmetry transformations of the form \(xtransform) have been recently
studied \[RamosRoca] in relation with the geometry of ${\ssf W}$ symmetry,
with the result that infinitesimal deformations of two-dimensional
curves have the algebraic structure of the standard
classical limit of Zamolodchikov's ${\ssf W}_3$ algebra.

We can also study how the coordinates $(t,\rho)$ of a given space-time 
point are changed after we perform this deformation.
Playing with the definition of the change of variables \(change) and
the relations \(v&ndef) and \(Freneteqns) for the tangent and normal
vectors one quickly arrives to
$$
\eqnalign{
\delta t=& -{\alpha\over e}-{\rho\over\Delta}{{\dot \beta}\over e^2},\cr
\delta \rho=&-\beta.\(coordchange)}
$$

The transformation for $\phi(t,\rho)$ is then simply a scalar field
transformation induced by the change of coordinates \(coordchange)
$$
\delta\phi(t,\rho)=-\delta t\;\partial_t\phi
-\delta\rho\;\partial_\rho\phi
=\left({\alpha\over e}+{\dot\beta\;\rho\over e^2\Delta}\right)
\partial_t\phi+\beta\,\partial_\rho\phi.
\(phitransform)
$$

It is obvious that the action $S_0[x,\phi]$ in \(curvedaction) is 
invariant under 
the gauge transformations generated by \(xtransform) and \(phitransform)
since it is just a rephrasal of $S_0[\varphi]$, the original
field-theoretical action, which is insensitive to any of these
transformations.

\medskip

Imagine now that one would be able to find an exact solution
$\phi[x(t)]$ of the equation of motion \(wrongeq).
Then, following the standard treament, upon inserting this solution 
back into the action we should get an exact expression $S_{\rm
eff}[x]$ for
the effective action. But, as we said earlier, the very same expression
should be obtained when using an arbitrary deformation of $x^\mu(t)$.
So we have to conclude that the (exact) effective action that one
would obtain in this way can be nothing but a trivial
($x(t)$-independent) one or, at most, a ``topological'' one giving
rise to no dynamics at all for $x^\mu(t)$.

We can give a more explicit proof of this with the following simple
argument.
Consider the two Noether identities corresponding to the gauge
transformations generated by $\alpha$ and $\beta$ in \(xtransform). 
Using deWitt condensed notation they read
$$
\eqnalign{
{\partial S_0\over \partial x^\mu}\; v^\mu+
{\partial S_0\over \partial \phi}\;\delta_\alpha\phi=&\;0,
\cr
{\partial S_0\over \partial x^\mu}\; n^\mu+ 
{\partial S_0\over\partial\phi}\;\delta_\beta\phi=&\;0.
\(noetherid)}
$$

Substituting any solution $\phi[x(t)]$ of the equation \(wrongeq), 
$\partial S_0/\partial\phi|_{x(t)}=0$, 
into these identities we get that $\partial 
S_0/\partial x$ is orthogonal to both $v$ and $n$, so it must be 
identically zero {\it off-shell}, \ie\ for any curve $x(t)$.
This implies directly that the would-be effective action, $S_{\rm
eff}[x]=S_0[x,\phi[x]]$, is in fact independent of $x(t)$!
$$
{\partial S_{\rm eff}\over\partial x}=
{\partial S_0\over\partial x}
+{\partial S_0\over\partial \phi}\;{\partial \phi[x]
\over\partial x}\equiv0.
\(trivialact)
$$

It should be clear by now what was the source of trouble in this
procedure. We were trying to find an effective action for the zero
modes of the field $\phi$ and describe them in terms of $x(t)$. But
we have introduced $x(t)$ without taking care to remove these degrees of
freedom from the field itself. The result was an overcounting of zero
modes that effectively rendered $x(t)$ spurious.

An obvious solution to this problem will be to add into the action a
Lagrange multiplier enforcing the constraint \(core).  If one proceeds
in this way the equation of motion \(wrongeq) gets modified by the
presence of a $\delta(\rho)$--type inhomogeneous term on the
right-hand-side. The consequence of this new contribution is to
produce a non-analytic behavior for the field $\phi(t,\rho)$ at
$\rho=0$, the location of the core.%
\fnote{Although from a somewhat different perspective,
Carter and Gregory \[CarterGregory] arrived also at the same
conclusion, {\it i.e.}~that the field $\phi$ need not satisfy the
equation of motion at $\rho=0$ and that a non-analytic behavior should
be expected at these points, when trying to find solutions of the
lowest order equation of motion \(phi1corr).}

Although this approach is technically and conceptually correct, one
may well feel uneasy about this sort of singularities which have been
originated by the choice of \(core), defining $x^\mu(t)$ to be the core
of the field $\phi$.  However, it is by no means mandatory that the
worldline describing the kink should necessarily be the core of
$\phi$.  This is because one has to assign a point-like trajectory
$x^\mu(t)$ to the extended object described by $\phi$ and there is
unavoidably some amount of freedom on how this can be done.  In fact,
an equally acceptable choice is given by the constraint
$$
\chi(t)=\int d\rho\;\kink'(\rho)\left(\phi(t,\rho)
-\kink(\rho)\right)=0,
\(gaugefix)
$$
which has been used earlier as the definition of the interface in
condensed matter physics \[Diehletal]\[Zia].  This is because
\(gaugefix) enforces $\delta\phi=\phi(t,\rho) -\kink(\rho)$ to have a
null component in the ``direction'' of the zero mode
$\kink'(\rho)$. Thus, it ensures that the dynamics of the zero mode is
no longer described in terms of the field $\phi(t,\rho)$ but in terms
of the world-line variables $x^\mu(t)$.  In addition, it is not
difficult to show that the curve $x^\mu(t)$ satisfying \(gaugefix)
actually coalesces with the core of the field for large values of the
kink's mass.  In this sense, \(gaugefix) can be regarded as a smoother
version of the constraint \(core).

Borrowing the language of gauge theories, we can say that \(gaugefix)
is just a gauge-fixing that we use in order to eliminate the gauge
freedom associated with deformations of the curve. Setting it will 
leave the reparametrizations of $x(t)$ as the only remnant gauge 
symmetry in the theory.

We need to make sure that this constraint places no actual physical
restrictions to the model.  We should then prove that \(gaugefix) is a
``good'' gauge-fixing, meaning that it can always be reached with the
help of a suitable gauge transformation. This can be rephrased as
follows.  Once a field configuration $\varphi(x^\mu)$ is given, there
should be (at least, locally) a unique world-line such that
\(gaugefix) is satisfied.

To show it, suppose that $x(t)$ is infinitesimally away from obeying the
constraint, so that we have
$$
\int d\rho\;\kink'(\rho)(\phi(t,\rho)-\kink(\rho))=\delta G(t).
\(almostthere)
$$
Then we can perform a general deformation $\delta x(t)$ of the form
\(xtransform) which,
according to \(coordchange), is tied with a coordinate change
$(t,\rho)\rightarrow (\tilde t,\tilde\rho)=(t+\delta
t,\rho+\delta\rho)$, that will take us there:
$$
\eqnalign{
\int d\tilde\rho\; & \kink'(\tilde\rho) \left(\tilde\phi(\tilde
t,\tilde\rho)-\kink(\tilde\rho)\right)=
\cr
=&\int d\rho\;\left[\;\kink'(\rho)-\kink''(\rho)\;
\beta(t)\;\right]
\left[\;\phi(t,\rho)-\kink(\rho)+\kink'(\rho)\;\beta(t)\;\right]
\cr
=&\;\delta G(t)\;+\;\beta(t)\;M_K\left(1-{1\over M_K}
\int d\rho\;\kink''(\rho)
\;\left(\phi(t,\rho)-\kink(\rho)\right)\right)=0.
\(goodfixing)}
$$
It might happen that the factor multiplying $\beta$ in \(goodfixing)
could vanish for some value of $t$. This will not be the case,
however, for configurations $\phi(t,\rho)$ departing very little from
$\kink(\rho)$ which is the regime we are considering throughout the
paper.  Therefore, we see from equation \(goodfixing) that $\beta(t)$
is determined uniquely in terms of $\delta G(t)$ and
$\phi(t,\rho)$. So we conclude that we can always find a (unique, up
to reparametrizations) world-line $\tilde x(t)=x(t) +\delta x(t)$ for
which the constraint \(gaugefix) is satisfied.

\medskip

We can impose \(gaugefix) with a Lagrange multiplier $g(s)$.
So the appropriate action to consider will be
$$
S[x,\phi,g]=S_0[x,\phi]+\int ds\;g(s)\chi(s),
\(gfaction)
$$
from which we shall get the effective action for $x(s)$.

The correct equation of motion for $\phi$ can now be obtained from
\(gfaction) and is given by
$$
{1\over\Delta}\;\partial_s\;{1\over\Delta}\;\partial_s\;\phi
\;-\;{1\over\Delta}\;\partial_\rho\;\Delta\;\partial_\rho\;\phi
\;+\;m^2(\phi^2-1)
\;\phi\;-\;{g(s)\over\Delta}\;\kink'(\rho)=0,
\(phieq)
$$
together with the constraint \(gaugefix). 

Equation \(phieq) differs from \(wrongeq) only in the last term, which
depends on the Lagrange multiplier $g(s)$. Its presence, however, has
important consequences for the behavior of the solutions found, both
quantitatively and qualitatively.

Indeed, it is direct to show that the kink-like configuration
$$
\phi(s,\rho)=\kink(\rho),\quad\quad
g(s)=-k(s),
\(kinklikesol)
$$
is not just an approximate solution to leading order, as it was for
the (non-fixed) equation \(wrongeq) in \S2.  It is actually an {\it
exact solution} of the new equation of motion \(phieq) and, as we
said, it corresponds to a kink at rest in a reference frame which is
co-moving with the world-line $x(s)$.  It gives rise to the (also
exact) effective action
$$
S_{\rm eff}[x]=-M_K\int ds,
\(freeaction2)
$$
whose only extrema are no other but the standard freely moving kink
solutions, \ie\ those satisfying $k=0$.

In view of this solution a few comments are in order.  First of all,
we have obtained it without having to impose any particular behavior
for the tangent and normal derivatives of $\phi$.  As we showed in the
previous section, this was an assumption generally made (see for
example ref.\ \[Gregory]) when trying to solve the equation \(wrongeq)
perturbatively.  Furthermore, for that equation the configuration
$\phi=\kink(\rho)$ was just an approximate solution, valid only to the
leading order in an $\epsilon$ expansion.  Because of that, one
expected, in addition to the basic mass term, higher order curvature
corrections to be generated as well.  On the contrary, the kink-like
solution \(kinklikesol) obtained here is an exact solution of the
properly fixed equation \(phieq) for a generic ``background''
$x(s)$. This implies that no curvature corrections need to be present
and that the pure mass term \(freeaction2) is already an exact
expression for the effective action.

\section{Perturbations around kink-like solutions}

We have shown in the previous section that $\phi=\kink(\rho)$ is a
solution of the equation of motion \(phieq), giving rise to no 
curvature terms in the effective action.
However, this does not exclude the possibility of having other solutions
for $\phi$ that could give rise to new effective actions.
This should not be a surprise because it is well-known \[ArodzWegrzyn]
that the effective action method leads to expressions for $S_{\rm
eff}$ which in general depend on the initial conditions imposed on
the fields that have been eliminated.
These different effective actions should be regarded as describing 
the system in different physical regimes, defined by the boundary 
or initial conditions that we set on the ``integrated'' field $\phi$.

The obvious way to seek new expressions for the effective action is to
perturb around the kink-like solution \(kinklikesol) and require
$\delta\phi$ to satisfy the linearized version of eq.\ \(phieq).

In this sense, it is reasonable to ask whether static (in the co-moving
frame) perturbations to \(kinklikesol) can actually exist.
If that were the case, we could expect the effective action to
contain, in addition to the standard mass term,
curvature-dependent corrections in $S_{\rm eff}[x]$ that would
account for the necessary energy increment in the co-moving frame.

Consider a perturbation around the solution \(kinklikesol)
$$
\eqnalign{
\phi(\rho)&=\kink(\rho)+\delta\phi(\rho),
\cr
g(s)&=-k(s)+\delta g(s),
\(klperturb)\cr}
$$
where we assumed $\delta\phi$ not to depend on the proper-time $s$.
The equation to solve, to the lowest order in $\delta\phi$ and $\delta
g$, is given by
$$
-{1\over\Delta}\;\partial_\rho\;\Delta\;\partial_\rho\;\delta\phi\;
+\;m^2(3\kink^2-1)\;\delta\phi\;-\;{\delta g\over\Delta}\;\kink'=0,
\(perturbeq2)
$$
and subject to the constraint
$$
\int d\rho\;\kink'\,\delta\phi=0.
\(staticconstr)
$$

We know from the discussion in \S 2 that all our formulation will
be correct only for the regime $\Delta\approx1$.
However, we keep for the moment an exact dependency on the curvature
$k(s)$ in order to account for a generic off-shell $x(s)$.

Using the orthonormal basis $\{|n\rangle\}$ defined by the eigenvalue
equation
$$
\left[-{d^2\over d\rho^2}
+m^2\left(3\kink^2(\rho)-1\right)\right]|n\rangle
=\omega^2_n\;|n\rangle,
\(eigeneq)
$$
we can express $\delta\phi$ in the form
$$
\delta\phi(\rho)=\sum_{n>0}\delta a_n\;|n\rangle.
\(modeexp)
$$
The index $n$ labels formally both the discrete and the continuous
part of the spectrum. The explicit expressions for $\omega_n$ and
$|n\rangle$ can be found in \[MorseFeshbach]\[Rajaraman], but we won't
need them here.

Notice that, being $|0\rangle\sim\kink'$ the zero mode, the constraint
\(staticconstr) is satisfied in \(modeexp) by simply excluding 
from the expansion the $n=0$ component.
The constant coefficients $\delta a_n$ should depend on the curvature
in a non-local way, \ie\ in terms of integrals of $k(s)$ along
the world-line $x(s)$.
By using the decomposition \(modeexp) we quickly obtain the formal 
expression for $\delta g(s)$,
$$
\delta g(s)\sim k(s)\sum_{r>0}\delta a_r\;
\langle0|(\omega^2_r\rho-{d\over d\rho})|r\rangle,
\(deltag)
$$
and a set of relations for the $\delta a_n$ coefficients,
$$
\delta a_n\;\omega_n^2+k(s)\sum_{r>0}\delta a_r\;
\langle n|(\omega^2_r\rho-{d\over d\rho})|r\rangle=0,
\(aneq)
$$
for $n>0$.
The only $s$-dependency in these relations shows up in the factor $k(s)$
in front of the second term. So, for example, deriving with respect
to $s$ will imply either $\delta a_n=0$, for all $n$, resulting in
$\delta\phi=\delta g=0$, or $\dot k=0$.
This second possibility, however, leads to the same conclusion. This
is because, being $k$ a constant in this case, we should be able to
write a power series expansion 
$$
\delta a_n=\sum_{p\geq0}a_n^{(p)}k^p
\(danexp)
$$
which, upon substitution in equation \(aneq) implies again $a^{(p)}_n=0$ 
for all $p$ and $n$.

\medskip

Another, simpler, argument showing that $\delta\phi(\rho)$ must vanish
goes as follows.
Consider the expression of the momentum $P_\mu$ as obtained from the
stress-energy tensor of the $\varphi^4$ theory \(action) and rewrite
it in the $(s,\rho)$ coordinates. Expanding around \(kinklikesol) we have
$$
\eqnalign{
P_\mu\;=\;&v_\mu(s)\left[M_K+\int d\rho\;\left({(\delta\dot\phi)^2\over
2\Delta^2}+{(\delta\phi')^2\over2}+{m^2\over2}(3\kink^2-1)
\;(\delta\phi)^2\right)+O((\delta\phi)^3)\right]
\cr
-&\;n_\mu(s)\int d\rho\;{\delta\dot\phi\over\Delta}(\kink'+
\delta\phi').
\(momentum)}
$$
By restricting ourselves to perturbations of the form
$\delta\phi(\rho)$ its expression will reduce to
$$
P_\mu=(M_K+\delta M)\;v_\mu(s),
\(staticp)
$$
where $\delta M$ does not depend on $s$.

Because of translational invariance, the equations of motion for
$S_{\rm eff}[x]$ are always given by $\dot P_\mu=0$, the dynamical
contents of the equation being hidden in the explicit expression
of $P_\mu$.
But it is clear that momentum conservation of \(staticp) leads to free
motion and that would be a globally static (in an appropriate frame)
solution with a rest mass given by $M_K+\delta M$.
However, we know that the standard kink solution, with a mass $M_K$,
is the only globally static solution in this topological sector.
So it can only be that $\delta M=0$ which in turn implies 
$\delta\phi(\rho)=0$.

\medskip

So we have to conclude that there are no modifications to the standard
mass term \(freeaction2) induced by perturbations of the form
\(klperturb).
It is clear that more generic ($s$-dependent) perturbations could also
be considered.
This situation, however, is very
likely to generate non-local expressions for the effective action 
$S_{\rm eff}[x]$ due to the explicit $s$-dependency induced by the 
solution obtained for $\delta\phi$.
Again, this is not a surprise since non-locality is also a known 
feature of the effective action method in a generic setting.
We address the interested reader to reference \[ArodzWegrzyn] by
Arod\'z and W\c egrzyn where a detailed study of such type of problems
can be found.

At any rate, an important point that we want to convey here is that
there is not a unique solution for this classical effective action.
This is a consequence of the different boundary conditions that can 
be imposed on the ``integrated'' field $\phi$ where each set of 
conditions will correspond to a different regime of the system.
This issue was somehow obscured in the treatment described in \S2 
where very specific corrections to $\kink(\rho)$, governed to lowest
order by eq.\ \(phi1corr), seemed to be necessarily present.
As we showed in \S3 this is not really the case because
$\kink(\rho)$ is in fact an exact solution. Then such perturbations 
actually satisfy an homogeneous equation and may or may not be
present depending on the boundary conditions that are set on them.

\section{Final comments}

We hope to have provided enough evidence that it is necessary to
properly dispose of the redundancy introduced by the change of
variables \(change) in order to get meaningful results for the 
effective dynamics of zero modes.

We have focussed in this paper on the simple two-dimensional kink
for the sake of illustration, finding that the basic mass term is
already an exact solution and that, in this case, no curvature 
corrections are present. 
However, it is clear that the analysis carried out here will apply as 
well in physically more interesting problems such as domain walls in
higher dimensions, where the issue of curvature corrections can 
also be addressed. Here we shall only sketch the procedure.

$D$-dimensional domain walls come out as solutions of the field
theory \(action) in $(D+1)$ dimensions.
The analog of the coordinate change \(change) will be in this case,
$$
x^\mu=x^\mu(\sigma)+\rho \;n^\mu(\sigma),
\(dwchange)
$$
where $x^\mu(\sigma^\alpha)$ parametrizes the embedding of the domain
wall in space-time. It gives rise, after a gauge-fixing analogous to 
\(gaugefix), to an equation for the field $\phi(\sigma,\rho)$ which 
is quite similar to \(phieq).
However, it can be shown that the configuration $\phi=\kink(\rho)$ is
not an exact solution here, for
arbitrary values of the extrinsic curvature.
Fortunately, this can be solved by changing the gauge-fixing
from the one used for the kink \(gaugefix) to a new one given by
$$
\chi(\sigma)=\int d\rho\;\Delta'\;\kink'(\rho)\left(\phi(\sigma,\rho)
-\kink(\rho)\right)=0.
\(gaugefixdw)
$$
where 
$$
\Delta=1+\rho K+{\rho^2\over2} R+\ldots,
\(dwdelta)
$$
is the higher dimensional
analog of \(Deltadef).
Changing the gauge-fixing amounts to picking up a slightly different
space-time embedding $x(\sigma)$, to represent the location of the
wall, once a field configuration $\varphi(x^\mu)$ is given.
It can be shown that this new gauge choice makes $\phi=\kink(\rho)$ an
exact solution of the $\phi$ equation of motion. This implies that the
standard Nambu-Goto action, together with extrinsic curvature 
contributions arising from \(dwdelta), will also be an
exact expression for the effective action.
Now, just as we did for the kink in \S4, it would be interesting to
perturb this solution and check whether new curvature contributions can
arise in this case.

\vskip 0.5truecm

\ack
We would like to thank Eduardo Ramos for whole-hearted discussions on
the subject and for a critical reading of the manuscript.

This work has been supported in part by a CICYT contract AEN95-0590,
by Ministerio de Educaci{\'o}n y Ciencia of Spain and by Generalitat de
Catalunya.

\vskip 0.5truecm

\refsout
\bye